\documentclass[reprint,
 amsmath,amssymb,
 aps,
showkeys,
]{revtex4-1}
\usepackage{enumerate}
\usepackage{siunitx}
\usepackage{graphicx}
\usepackage{dcolumn}
\usepackage{braket}
\usepackage{bm}
\usepackage{color}
\usepackage{xr}
\usepackage{lineno}

\begin{document}

\title{Magnon hydrodynamics in an atomically-thin ferromagnet}

\author{Ruolan Xue$^{1*\dag}$, Nikola Maksimovic$^{2*\dag}$, Pavel E. Dolgirev$^2$, Li-Qiao Xia$^3$, Aaron M{\"u}ller$^4$, Ryota Kitagawa$^5$, Francisco Machado$^{2,6}$, Dahlia R. Klein$^{3,7}$, David MacNeill$^3$, Kenji Watanabe$^8$, Takashi Taniguchi$^8$, Pablo Jarillo-Herrero$^3$, Mikhail D. Lukin$^2$, Eugene Demler$^4$, Amir Yacoby$^{1,2}$}

\email{ruolanxue@g.harvard.edu
\\
nikola$_$maksimovic@fas.harvard.edu
\\yacoby@g.harvard.edu}

\affiliation{$^1$John A. Paulson School of Engineering and Applied Sciences, Harvard University, Cambridge, MA, USA \\
$^2$Department of Physics, Harvard University, Cambridge, MA, USA \\
$^3$Department of Physics, Massachusetts Institute of Technology, Cambridge, MA, USA
\\
$^4$Department of Physics, Eidgenössische Technische Hochschule Zürich, Zurich, Switzerland
\\
$^5$Department of Electrical and 
Electronic Engineering, Tokyo Institute of Technology, Tokyo, Japan
\\
$^6$ITAMP, Harvard-Smithsonian Center for Astrophysics, Cambridge, MA, USA
\\
$^7$Department of Condensed Matter Physics, Weizmann Institute of Science, Rehovot, Israel
\\
$^8$Research Center for Functional Materials, National Institute for Materials Science, Tsukuba, Japan}
\thanks{These authors contributed equally to this work.}

\date{\today}

\begin{abstract}

Strong interactions between particles can lead to emergent collective excitations. These phenomena have been extensively established in electronic systems, but are also expected to occur for gases of neutral particles like magnons, i.e. spin waves, in magnets. In a hydrodynamic regime where magnons are strongly interacting, they can form a slow collective density mode -- in analogy to sound waves in water -- with characteristic low-frequency signatures. While such a mode has been predicted in theory, its signatures have yet to be observed experimentally. In this work, we isolate exfoliated sheets of CrCl$_3$ where magnon interactions are strong, and develop a technique to measure its collective magnon dynamics via the quantum coherence of nearby Nitrogen-Vacancy (NV) centers in diamond. We find that the thermal magnetic fluctuations generated by monolayer CrCl$_3$ exhibit an anomalous temperature dependence, whereby fluctuations increase upon decreasing temperature. Our analysis suggests that this anomalous trend is a consequence of the damping rate of a low-energy magnon sound mode which sharpens as magnon interactions increase with increasing temperature. By measuring the magnetic fluctuations emitted by thin multilayer CrCl$_{3}$ in the presence of a variable-frequency drive field, we observe spectroscopic evidence for this two-dimensional magnon sound mode.

\end{abstract}

\maketitle

\section{Introduction}

Modern quantum technologies are often based on magnetic materials, with applications ranging from non-boolean computing~\cite{csaba2014spin}, to spin torque manipulation~\cite{locatelli2014spin}, to overcoming the Landauer limit of energy consumption~\cite{berut2012experimental,cuykendall1987reversible}. These technologies rely on the propagation of magnons -- quanta of spin excitation --- in magnetic materials~\cite{chumak2015magnon}. A fundamental understanding of various magnon transport mechanisms is therefore crucial for the development, design, and operation of spintronic devices, especially those based on recently-discovered atomically-thin magnets~\cite{burch2018magnetism}.

Of particular interest is the realization of novel transport regimes. Typically, propagating magnons lose their momentum through collisions with the boundaries of the sample or the atomic lattice of the host material. A qualitatively different regime emerges when magnons collide predominantly with each other. Such a hydrodynamic regime is characterized by liquid-like flow in which transport is controlled by collisions within the fluid rather than its interactions with the host material. This hydrodynamic regime in a gas of magnons is expected to exhibit a collective sound mode at energies lower than that of single magnon excitations in analogy to second-sound in superfluids~\cite{michel1970hydrodynamic,reiter1968magnon,rodriguez2022probing,hohenberg1965microscopic}. 

In the solid state, hydrodynamic transport is an emerging field with growing evidence in electron systems~\cite{sulpizio2019visualizing,levitov2016electron,bandurin2016negative,crossno2016observation,moll2016evidence}, and even demonstrated practical value~\cite{geurs2020rectification}. In theory, it has been suggested that a hydrodynamic regime could be realized in densely-populated and strongly interacting magnon gases, which exist in highly isotropic ferromagnets~\cite{schwabl1970hydrodynamics,rodriguez2022probing,dyson1956general,reiter1968magnon,ulloa2019nonlocal}. CrCl$_3$, a layered magnetic insulator with relatively low single-ion anisotropy~\cite{ narath1963low,kostryukova1972specific,mcguire2017magnetic,kim2019evolution,kim2019giant,macneill2019gigahertz}, is therefore an excellent candidate. This material can further be isolated in atomically-thin form~\cite{bedoya2021intrinsic,cai2019atomically,klein2019enhancement}. However, it is challenging to measure collective dynamics of magnons in general, and even more so in atomically-thin materials because of their mesoscopic sample volumes and correspondingly small signals.

These difficulties could be naturally overcome using atomic-like defects such as nitrogen-vacancy (NV) centers in diamond, which are capable of detecting small magnetic field signals at the nanoscale~\cite{schirhagl2014nitrogen,degen2017quantum}. Previous studies demonstrated how proximal NV centers can be used to measure the dissipative magnetic susceptibility of a material through noise measurements in the GHz frequency range (also called $T_{1}$ relaxometry), via the fluctuation-dissipation theorem~\cite{wang2022noninvasive,mclaughlin2022quantum,flebus2018quantum}. Here, we employ a magnetic noise sensing technique based on the dephasing rates ($T_{2}$) of NV centers in proximity to atomically-thin CrCl$_{3}$. We find that the low-frequency magnetic noise generated by a CrCl$_{3}$ monolayer exhibits an enhancement with decreasing temperature. This observation is surprising given that thermal fluctuations generated by an ordered magnet are expected to freeze as temperature is reduced~\cite{slichter2013principles}. We suggest that the observed trend is a signature of a low-energy hydrodynamic magnon sound mode with anomalous damping rate set by magnon viscosity.

In order to probe the magnon sound mode directly, we use the NV centers to measure the magnetic response of a thin bulk CrCl$_{3}$ sample in the presence of a variable-frequency magnetic drive, revealing direct spectroscopic evidence for a magnon sound wave within the ferromagnetic layers of CrCl$_{3}$.

\begin{figure*}[htbp!]
\centering
\includegraphics[width=0.7\textwidth]{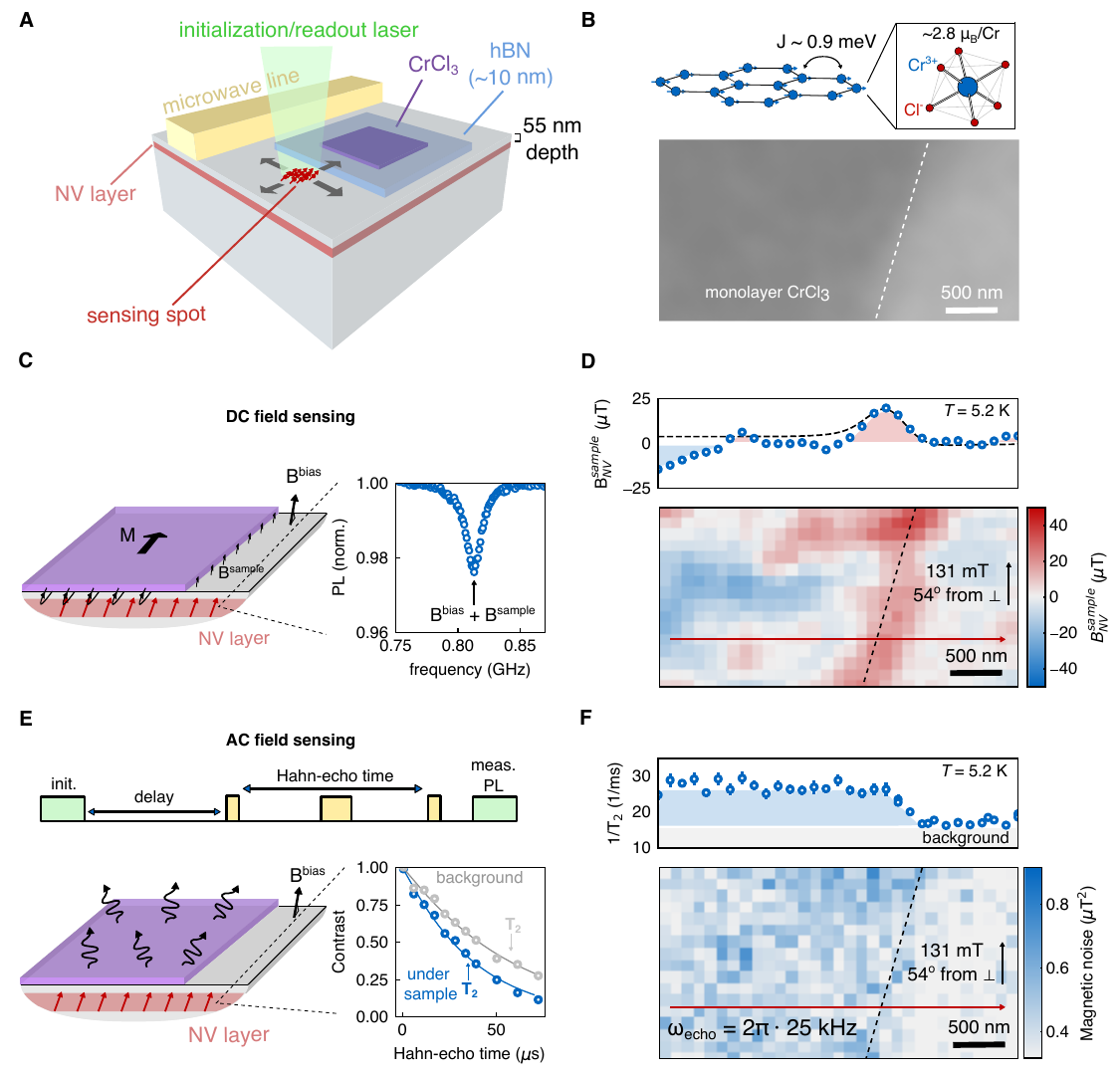}
\caption{CrCl$_{3}$ material properties and nitrogen-vacancy (NV) center sensing techniques.~\textbf{A} CrCl$_{3}$ samples are embedded in a coplanar waveguide on the surface of a diamond slab, which contains a uniform layer of shallow NV centers. A focused laser addresses a movable sensing spot in the NV layer. 
~\textbf{B} Crystal structure of the monolayer CrCl$_{3}$ and its relevant magnetic exchange interaction along with an optical image of a region of a monolayer CrCl$_{3}$.~\textbf{C} Static magnetic field, with contributions from both the sample's stray field and the DC bias field, is measured via optically-detected magnetic resonance.~\textbf{D} $B^{\rm sample}_{\rm NV}$ is isolated by rastering the sensing spot position across the sample and subtracting the bias field determined off the sample. The stray field profile across a linecut (upper panel) is consistent with a simulation using a sample magnetization of 12 $\mu_{B}$/nm$^{2}$, 85$^{o}$ from the surface normal (dashed line).~\textbf{E} AC field sensing is achieved by measuring the dephasing of the NV centers.~\textbf{F} An image of the magnetic noise intensity measured using a fixed Hahn-echo time of 40$\mu$s ($\omega_{echo} = 2\pi \cdot 25$ kHz), taken under the same conditions as in~\textbf{D}. The line cut shows $1/T_{2}$ as obtained from a full fit of the Hahn-echo decay at each spatial point.
}
\label{fig1}
\end{figure*}

\section{Sensing techniques}

Exfoliated CrCl$_{3}$ samples encapsulated by hBN are placed in the gap of a coplanar waveguide patterned on the surface of a 2$\times$2$\times$0.5 mm bulk diamond. The waveguide is used to supply microwaves to drive the spin state of NV centers implanted on average 55$\,$nm beneath the surface of the diamond (Fig.~\ref{fig1}A). A sensing spot of about 640 nm diameter is selected within the NV layer using a focused green laser that can be spatially rastered around the CrCl$_{3}$ sample.

In CrCl$_{3}$, the Cr atoms form a honeycomb lattice and are characterized by nearest-neighbor ferromagnetic exchange interaction (Fig.~\ref{fig1}B)~\cite{lu2020meron,mcguire2017magnetic}. To probe the static magnetic properties of the material, we utilize the shift in the NV's electron spin resonance frequency (see Fig.~\ref{fig1}C and SI II.A) to measure the stray field generated by the edge of the sample (Fig.~\ref{fig1}D). 

Access to magnetization dynamics is achieved through a quantum coherent sensing protocol.
The Hahn-echo sequence leverages phase-coherent pulses to evolve the NV spin state on the Bloch sphere, after which the spin contrast is recorded (Fig.~\ref{fig1}E). As the sequence time is continuously varied, environmental magnetic noise dephases the NV spin, exponentially suppressing the spin contrast with a characteristic rate $1/T_{2}$ (Fig.~\ref{fig1}E inset).
We isolate the sample-induced dephasing rate by subtracting the background rate obtained on nearby bare diamond (Fig.~\ref{fig1}F and SI II.C). To ensure that the signal obtained reflects the equilibrium noise, a variable delay between 1 $\mu$s and 60 $\mu$s was added between the initialization laser pulse and the Hahn-echo sequence~\cite{zhang2023enhanced} (see SI V). 

The sample-induced magnetic noise in this measurement scheme is proportional to the dissipative component of the magnetic susceptibility of the material $\text{Im}[\chi(k,\omega)]$, subject to frequency and momentum filtering functions~\cite{machado2023quantum}. The probed frequency $\omega_{\rm echo}$ is essentially set by the precession time of the Hahn-echo sequence, and the probed momentum $k$ is strongly peaked at momenta given by the inverse of the NV-sample distance $z_{\rm NV}$. The $T_{2}$-signal is thus related to $\text{Im}[\chi(k_{\rm NV},\omega_{\rm echo})]$~\cite{machado2023quantum} (SI X and V): 
\begin{equation}
    \frac{1}{TT_{2}} \propto
    \frac{\text{Im}[\chi(k_{\rm NV},\omega_{\rm echo})]}{\omega_{\rm echo}},
    \label{eq:fluctuation-dissipation_v2}
\end{equation}
where $T$ is temperature.
In this experiment, $k_{\rm NV} \approx 1/z_{\rm NV}$ with $z_{\rm NV} = 65\,$nm and $\omega_{\rm echo}\sim\, 2\pi \cdot (10-1000)\,$kHz. Access to these relatively short length-scales distinguishes NV noise measurements from conventional low-frequency probes of dynamical susceptibility~\cite{topping2018ac}.

\section{Results}

\begin{figure}[hbtp!]
\centering
\includegraphics[width=0.47\textwidth]{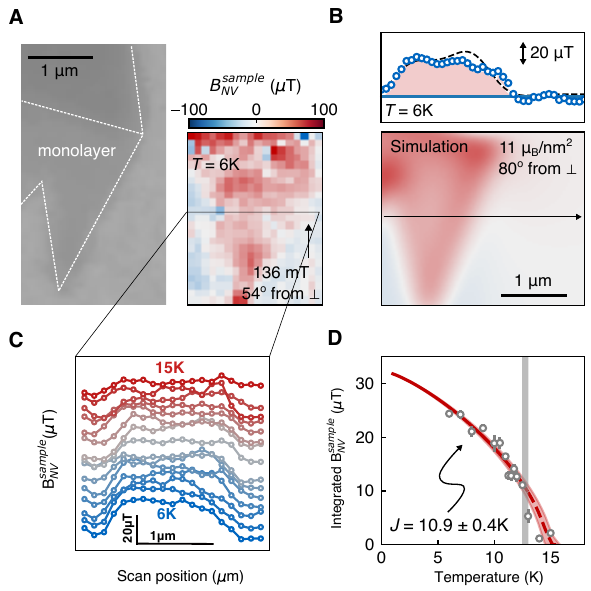}
\caption{Static magnetization properties of a monolayer CrCl$_{3}$.~\textbf{A} Optical image alongside a stray field map taken with a 136$\,$mT bias field applied 54$^{\circ}$ from the surface normal at 6$\,$K.~\textbf{B} Best-fit simulated stray field profile plotted in the same color scale as the stray field map in (A), corresponding to a sample magnetization of 11 $\mu_{B}$/nm$^{2}$ 80$^{o}$ from the surface normal. The upper panel depicts a single line scan (blue points) along with the simulation (dashed line).~\textbf{C} Temperature-dependent stray field profile over the waist of the sample.~\textbf{D} Integrated stray field over the area of the sample as a function of temperature with a grey bar denoting the ferromagnetic-to-paramagnetic transition temperature. The red line is a fit to our dipolar spin-wave model (SI XI), from which we determine the ferromagnetic exchange constant, $J$.}
\label{fig2}
\end{figure}

\subsection{Monolayer CrCl$_{3}$}

We first employ stray field measurements~\cite{bedoya2021intrinsic,rondin2014magnetometry} to estimate the ferromagnetic transition temperature and exchange constant of a monolayer CrCl$_3$ film. The magnetic stray field at 6$\,$K (Fig.~\ref{fig2}A) is compared to simulations of the sample geometry incorporating a variable magnetization magnitude and orientation (Fig.~\ref{fig2}B). This analysis yields a best fit solution corresponding to a nearly planar magnetization of 11 $\mu_B$/nm$^{2}$ (see also SI IV.A). The temperature-dependence of the integrated stray field (Fig.~\ref{fig2}C) reveals a continuous decrease in the magnetization with increasing temperature and a ferromagnetic-to-paramagnetic phase transition at $T_c \approx 12.5\,$K (Fig.~\ref{fig2}D), a value consistent with previous measurements on epitaxially-grown monolayer CrCl$_3$~\cite{bedoya2021intrinsic}.

We extract the exchange interaction $J$ by fitting the temperature-dependent stray field data within the ferromagnetic state to a self-consistent linearized spin-wave theory with two free parameters (details in SI IV and VII): $J$ and a scale factor converting magnetization to stray field (Fig.~\ref{fig2}D). Deviations between theory and experiment above 12.5$\,$K arise from the breakdown of the self-consistent linearized spin-wave theory near the transition temperature~\cite{bruno1991spin}. The resulting exchange constant ($J = 10.9 \pm 0.4 \,\rm K$) agrees well with previous measurements on multilayer CrCl$_{3}$~\cite{kim2019evolution}. The fitted scale factor converting magnetization to integrated stray field, assuming a zero-temperature magnetization of 2.8 $\mu_{B}$/Cr~\cite{bedoya2021intrinsic}, is within 25\% of that obtained via our best-fit simulations of the stray field profile at 6\,K (potential sources of systematic error in our best-fit simulation procedure are discussed in SI IV.A.)~\cite{broadway2020improved}. The overall agreement between our spin-wave model and stray field data indicates that dipolar interactions stabilize the ferromagnetic phase (see SI IX), and that magnons are the relevant excitations below 12.5$\,$K for modeling the magnetic dynamics of monolayer CrCl$_3$.

\begin{figure*}[htbp!]
\centering
\includegraphics[width=0.7
\textwidth]{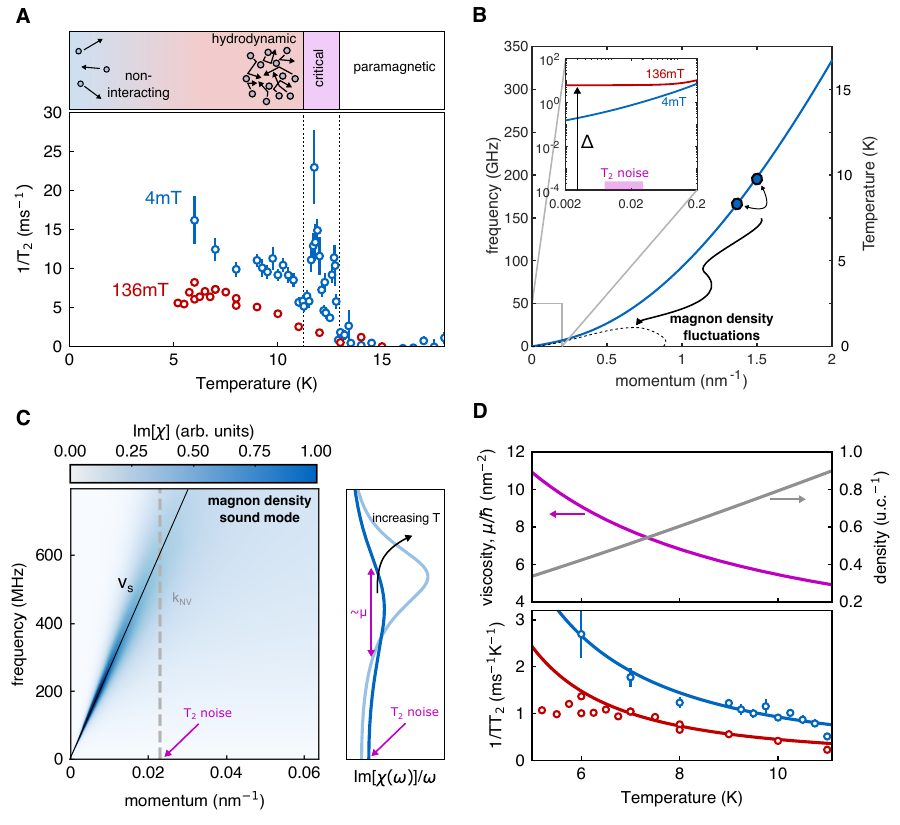}
\caption{Dephasing spectroscopy of a monolayer CrCl$_{3}$ and relation to magnon dynamics.~\textbf{A} Sample-induced $1/T_{2}$ dephasing rate, averaged over the area of the monolayer sample, at both 4$\,$mT external field normal to the sample and 136$\,$mT tilted 54$^{\rm o}$ from the normal.~\textbf{B} Calculated magnon dispersion along the $\Gamma$-K$'$ direction, at both applied magnetic fields as in the experiment. The inset shows a zoom-in with a label of the relevant frequency and momenta of the $T_{2}$ noise measurements (pink square). The dispersion is essentially parabolic with an effective gap $\Delta$ set by anisotropy terms. The right axis is the energy scale plotted in the unit of temperature.~\textbf{C} Im[$\chi$] of the magnon sound mode in a hydrodynamic regime. A linecut at $k_{\rm NV}$ depicts how $T_{2}$ noise (related to Im[$\chi$]/$\omega$ via Eq.~\eqref{eq:fluctuation-dissipation_v2}) is proportional to the viscous damping rate, $\mu$, of the sound mode.~\textbf{D} Top panel shows
calculated magnon density $n(T)$ and exchange-driven magnon viscosity $\mu(T)$, which decreases with increasing $T$.
The bottom panel shows a fit to the data between 6K and 11K based on our phenomenological theory in Eq.~\eqref{eq:viscosity_fit}.}
\label{fig3}
\end{figure*}

While the static magnetic properties conform to a spin-wave model~\cite{bedoya2021intrinsic,kim2019evolution}, the observed spin dynamics, as measured by $1/T_{2}$, unveil peculiar behavior indicative of the existence of strong magnon interactions within the ferromagnetic state. These measurements are performed both with a 4$\,$mT bias field normal to the sample plane and a 136$\,$mT field along the NV axis, 54$^{\circ}$ from normal (Fig.~\ref{fig3}A).

At 4$\,$mT, the noise exhibits sharp peaks close to the transition temperature as inferred from static field measurements. Subsequently, the noise level continues to rise with further cooling down to our lowest recorded temperature (Fig.~\ref{fig3}A). A 136$\,$mT magnetic field suppresses the features near the transition temperature, but only weakly affects the increasing noise trend with lowering temperature. 

The peak in noise near the phase transition confined between 11.5K and 13K likely results from critical fluctuations~\cite{machado2023quantum}. The suppression of this feature for the stronger magnetic field could result from the explicit symmetry-breaking effect of the planar component of the applied field. The presence of two closely-spaced peaks appears to be non-generic (see SI VI for measurements on another sample), pointing to spatial inhomogeneity in transition temperatures of a few percent on length scales of the sensing spot. Similar variation in transition temperatures has been observed previously in epitaxial CrCl$_{3}$ monolayers~\cite{bedoya2021intrinsic}.

The increase of $1/T_{2}$ with decreasing temperature within the ordered state ($<11$K) is a surprising observation given that, on general grounds, the magnetic noise in an ordered ferromagnet should monotonically decrease with decreasing $T$ as thermal fluctuations become progressively more frozen~\cite{slichter2013principles}. We consider a variety of mechanisms that could potentially produce noise in this material, including the damping of single magnons~\cite{sparks1961ferromagnetic,jermain2017increased}, magnetic vortices~\cite{lu2020meron}, magnetization tunneling~\cite{vitale1994thermal}, domain walls, exotic magnetic particles~\cite{dusad2019magnetic}, and spin glass physics~\cite{reim1986magnetic,quilliam2008evidence}, which all fail to consistently capture the experimental phenomenology (see SI XV.A). 

We argue that this anomalous noise scaling can be captured by magnon hydrodynamics. At the temperatures of the experiment, the magnons comprise a dense gas, which produces noise due to the transport of magnon density fluctuations through the NV's nanoscale sensing region~\cite{flebus2018quantum}. The magnons are so strongly-interacting in the monolayer CrCl$_{3}$ that the gas can form a low-energy long-lived collective sound mode, which both moves faster and becomes longer lived as temperature increases. This interaction-driven speed-up and linewidth narrowing counteracts the amplification of thermal fluctuations, ultimately leading to a suppression of noise with increasing temperature.

We first note that the $1/T_{2}$ signal is expected to be insensitive to creation and annihilation processes of single magnon excitations, i.e., transverse oscillations of the spins. To see this explicitly, we plot the magnon dispersion in monolayer CrCl$_{3}$ computed based on our linearized spin-wave theory in Fig.~\ref{fig3}B. 
At the typical experimental temperatures, the average magnon thermal energy is large compared to the anisotropy energy scales (dipolar, single-ion, and planar magnetic field). As such, thermal magnons essentially obey a parabolic dispersion with mass $m$ set by the exchange interaction and an effective small pseudo-gap $\Delta$, which originates from the weak effects of the anisotropy (Fig.~\ref{fig3}B, Fig. S6). 
The NV center filters out momenta away from $k_{\rm NV} \simeq 1/z_{\rm NV}$. Magnons near $k_{\rm NV}$ have frequencies $\gtrsim$ 1 GHz, which are highly off-resonant with the frequencies of the $1/T_{2}$ measurements (see inset of Fig.~\ref{fig3}B).

Instead, the $1/T_{2}$ signal originates from subgap density fluctuations of the magnon gas, i.e., longitudinal spin fluctuations, which originate from the scattering of high-momentum thermal magnons~\cite{flebus2018quantum, rodriguez2022probing}. A hydrodynamic regime can be realized when momentum-conserving magnon-magnon scattering dominates over relaxational mechanisms -- such as magnon-phonon, magnon-Umklapp, and magnon damping -- over the 65 nm length-scale probed by the NV. Our estimates (see SI XV.A) suggest that this is indeed the case in CrCl$_{3}$ over the temperature range measured in the experiment. The magnon-magnon mean-free-path given by $l_{\rm mfp}\simeq (J/T)^{3/2}/(n a)$ ~\cite{dyson1956general}, ranges between 18 nm at $T = 5\,$K and 3 nm at $T = 10\,$K, where $a$ is the lattice constant and $n$ is the thermal magnon density. By contrast, the relaxational length scales are all found to exceed 100 nm over the same temperature range (SI XV.A).

In the hydrodynamic regime, the longitudinal spin fluctuations form a low-energy collective sound mode whose broadening due to finite lifetime spans frequencies down to those of the $1/T_{2}$ measurements (Fig.~\ref{fig3}C). In particular, the sound mode has a linear dispersion set by the speed of sound $v_s$ and decay rate set by the magnon viscosity $\mu$. In this regime, the low-frequency susceptibility probed by the noise measurement is determined by the transport coefficients of the sound mode: $(TT_{2})^{-1} \propto \mu/{(m^{2}v_{s}^{4})}$, and the effective mass of the magnons, $m$. The exchange-collision induced magnon viscosity $\mu \propto q_{\rm th} n l_{\rm mfp} \sim 1/T$, where $n \sim T$, $q_{\rm th} \sim T^{1/2}$, and $l_{\rm mfp} \sim T^{-5/2}$ (see SI XIV.A). The temperature-dependent viscosity can therefore be captured by $\mu_k/T$ with $\mu_{k}$ a constant. For the speed of sound, in the high magnon density regime we operate in with $T \gg \Delta$,
we find that the temperature-dependent sound velocity is set by the magnon pseudogap $\Delta$: $v_s^2 \approx v^{2}_{0} |\log(\Delta/T)|$ (see SI XIII.B), with $v_{0}$ a constant. Taking into account these effects, the resulting response function scales with temperature as:

\begin{equation}
    \frac{1}{TT_{2}} \simeq \left(\frac{C}{z_{\rm NV}^{4}}\frac{\mu_{k}}{m^{2}v_{0}^{4}}\right) \frac{1}{T\log^{2}\left(\Delta/T\right)}.
    \label{eq:viscosity_fit}
\end{equation}

Here $C$ is a constant that encodes geometrical factors and the gyromagnetic ratio of the NV (see SI VI.B). This phenomenological theory produces excellent qualitative agreement with our 1/$T_2$ measurements between 6\,K and 11\,K (Fig.~\ref{fig3}D). Deviations at the lowest temperatures could arise as the mean free path approaches or exceeds $z_{\rm NV}$ and the system enters a non-interacting regime. The agreement further allows us to extract physical parameters of the system via a two-parameter fit (see SI VI.D). We obtain $\frac{C}{z_{\rm NV}^{4}}\frac{\mu_{k}}{m^{2}v_{0}^{4}} = 43 \pm 9$ kHz and $\Delta = 1.2 \pm 0.2$K at 4mT; $\frac{C}{z_{\rm NV}^{4}}\frac{\mu_{k}}{m^{2}v_{0}^{4}} = 41 \pm 10$ kHz and $\Delta = 1.7 \pm 0.7$K at 136mT. 
The roughly factor of two increase in $\Delta$ upon increasing magnetic field from 4 mT to 136 mT is in qualitative agreement with the results of our linearized spin-wave theory, which predicts $\Delta\simeq 0.17\,$K at $B = 4\,$mT and $\Delta\simeq 0.33\,$K at $B = 135\,$mT (see SI VII.C and Fig. S6). Quantitative disagreement here may result from magnon band renormalization effects which are not taken into account in the present theory~\cite{goldhirsch1980magnon}, and addressed further in the discussion. Our hydrodynamic model of the thermal magnon fluid, and in particular the $1/T$ scaling of the viscous damping rate, provides a well-motivated and satisfactory interpretation of the trend seen in the experiment. This interpretation is further supported by direct spectroscopic signatures of the hydrodynamic sound mode provided below.

\begin{figure}[htbp!]
\centering
\includegraphics[width=0.5
\textwidth]{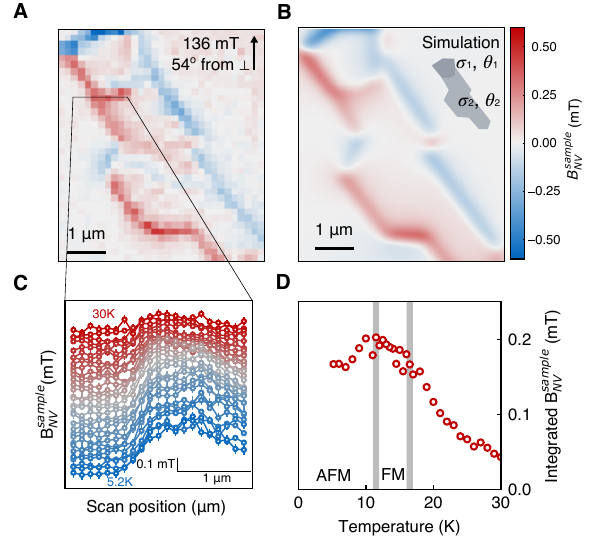}
\caption{Static magnetization properties of thin multilayer CrCl$_{3}$~\textbf{A} Stray field measurements at 6$\,$K with a 131$\,$mT field applied 54$^{\circ}$ from normal to the plane.~\textbf{B} Best-fit simulation with optimized magnetization $\sigma_1=29\,\mu_B$/nm$^2$ at 60$^\circ$ from surface normal (29\,nm thickness) and $\sigma_2=22\,\mu_B$/nm$^2$at 83$^\circ$ from surface normal (25\,nm thickness).~\textbf{C} Temperature-dependent stray field measured along a horizontal linecut on the sample.~\textbf{D} Integrated stray field versus temperature. Two features align with the expected interlayer AFM and intralayer FM ordering transition temperatures.}
\label{fig4}
\end{figure}

\subsection{Bulk CrCl$_{3}$}

\begin{figure*}[htbp!]
\centering
\includegraphics[width=0.75
\textwidth]{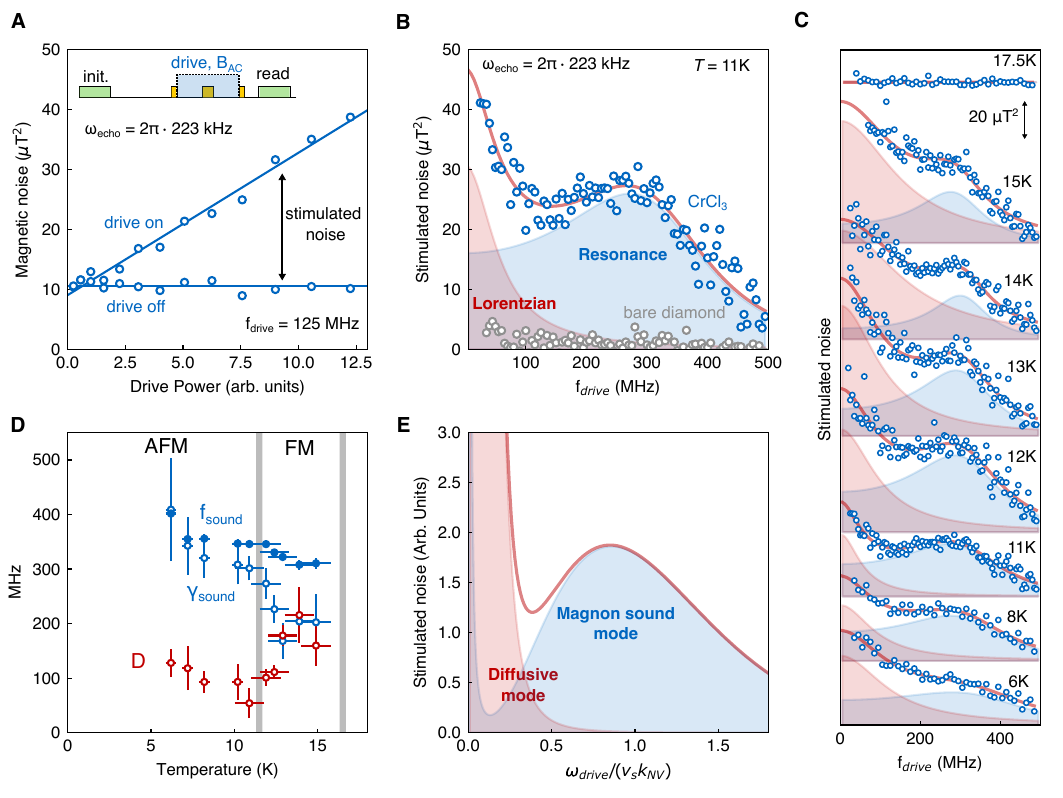}
\caption{Spectroscopic evidence for a hydrodynamic magnon sound mode in thin bulk CrCl$_{3}$~\textbf{A} Magnetic noise detected by a Hahn-echo sequence measured on the multilayer sample at 11\,K and static 1.3\,mT out of plane bias field. An applied oscillating magnetic drive field with frequency $f_{drive}$ is delivered by the coplanar waveguide during the measurement. The stimulated magnetic noise generated by the sample in this linear-response regime can be related to the magnetic response function of the sample at $f_{drive}$ and the NV momentum (see supplement XVIII).~\textbf{B} A spectrum measured by NV centers under the CrCl$_{3}$ sample (blue points), which is obtained by varying $f_{drive}$ at fixed power. The spectrum is well-fitted by the sum of a magnon sound mode resonance at finite frequency (blue) and a zero-frequency Lorentzian (red). Measurements on nearby bare diamond yield essentially no response.~\textbf{C} Temperature-dependent spectra with fits.~\textbf{D} The fitted resonance frequency, $f_{sound}$ (solid blue points), linewidth $\gamma_{sound}$ (open blue points), and Lorentzian width $D$ (open red points), as a function of temperature. Vertical grey bars indicate the inter- and intralayer transition temperatures of the material.~\textbf{E} A theoretical calculation of the stimulated noise spectrum arising from a magnon sound mode and a diffusive mode reproduces the lineshape seen in the experiment.}
\label{fig5}
\end{figure*}

We now focus on studying the hydrodynamic magnon sound mode in a CrCl$_{3}$ multilayer. This material is composed of ferromagnetic monolayers, which are weakly antiferromagnetically-coupled between layers~\cite{mcguire2017magnetic}. Previous literature suggests that the magnon dynamics are largely confined to the two-dimensional layers~\cite{pocs2020giant,chen2021massless}, suggesting that our model of a hydrodynamic mode could extend to this case -- indeed, multilayer CrCl$_{3}$ also exhibits an anomalous $1/T_{2}$ versus temperature, which to some extent mirrors the behavior observed in the monolayer (see SI VIII). 

We first characterize the static magnetic structure of the material. The measured static field of the CrCl$_{3}$ multilayer at 6\,K is consistent with layered antiferromagnetism, where the 131\,mT bias field cants the spins between layers by about 4-5\% (Figs.~\ref{fig4}A,B). This degree of canting is roughly consistent with typical saturation fields observed in exfoliated CrCl$_{3}$~\cite{wang2019determining}. By monitoring the stray field across a single linecut as temperature increases (Fig.~\ref{fig4}C), we observe two features in the temperature-dependent magnetization indicative of magnetic phase transitions (Fig.~\ref{fig4}D), above which the signal decays with increasing temperature characteristic of a paramagnetic response. Based on a comparison to existing literature on this material~\cite{mcguire2017magnetic,cai2019atomically,bykovetz2019critical}, we conclude that the material undergoes two phase transitions -- the first at $\sim$16.5K associated with ferromagnetic ordering within the layers, and the second at $\sim$12K associated with antiferomagnetic alignment between layers (Fig.~\ref{fig4}D). Measurements on a separate sample indicate the same qualitative behavior with slightly different transition temperatures (see SI VIII).

In order to access the $\sim$100MHz frequency response of this material where the intralayer hydrodynamic sound mode is expected to lie, we apply a magnetic field stimulus at frequency $f_{drive}$, and detect the material's response via a Hahn-echo sequence with the NV centers. In the linear response regime in which we operate (Fig.~\ref{fig5}A), the signal in this measurement can be theoretically related to the response function of the material at $f_{drive}$ subject to the NV momentum filtering function and the spin-echo frequency filtering function (see supplement XVIII).

By varying the drive frequency, $f_{drive}$, we detect a spectrum composed of a resonance at finite frequency and a broad low-frequency tail (Fig.~\ref{fig5}B). Motivated by a theoretical calculation of the anticipated experimental spectrum generated by a hydrodynamic sound mode and a diffusive component (see SI XVIII.D.), we employ a phenomenological fit to capture the temperature-dependent evolution of this spectrum (Fig.~\ref{fig5}C). In particular, the spectra can be fitted by a resonance with a frequency $f_{sound}$, linewidth $\gamma_{sound}$, and spectral weight $A_{sound}$, and a zero-frequency Lorentzian with a width $D$ and spectral weight $A_{lorentzian}$:
\begin{equation}
    \frac{A_{sound}}{(f^{2} - f^{2}_{sound})^{2} + f^{2}\gamma_{sound}^{2}} + \frac{A_{lorentzian}}{D^{2} + f^{2}}.
\end{equation}
The resonance frequency, $f_{sound}$, and linewidth, $\gamma_{sound}$, both exhibit smooth trends across the interlayer antiferromagnetic phase transition (Fig.~\ref{fig5}D), suggesting that this component of the spectrum reflects spin dynamics confined to the ferromagnetic layers which are largely insensitive to the interlayer alignment of spins. By contrast, the Lorentzian width, $D$, exhibits a more prominent feature across this phase transition (Fig.~\ref{fig5}D), becoming broader in the intralayer ferromagnetic phase than in the interlayer antiferromagnetic phase. This trend suggests that the Lorentzian spectral component contains contributions reflecting interlayer dynamics~\cite{flebus2018quantum}, but may also include a diffusive contribution generally present even in the hydrodynamic description, as discussed in SI XVIII.

It is first important to rule out the possibility that the resonance observed in Fig.~\ref{fig5}B arises from magnons. We note that the signal completely disappears above the ferromagnetic transition temperature (Fig.~\ref{fig5}C), and exhibits a weak field-dependence (see SI IX.F). These observations both strongly disagree with the expected behavior of a resonance caused by transverse spin oscillations, which should exhibit a strong dependence on external field and persist above the transition as paramagnetic resonance~\cite{macneill2019gigahertz}.

These facts indicate that the resonance observed in Fig.~\ref{fig5}B,C corresponds to a subgap hydrodynamic magnon sound mode within the 2D ferromagnetic layers of CrCl$_{3}$. Indeed, the shape of the experimental spectrum can be reproduced by our full theoretical model of the stimulated noise generated by the combination of a hydrodynamic mode and a phenomenological low-frequency diffusive mode (Fig.~\ref{fig5}E). This theoretical calculation implies that the peak position of the resonance can be used to roughly estimate the velocity of the sound mode: via $v_{s} \approx 2\pi f_{sound}/k_{\rm NV} \sim 150\,$m/s. This value, while representing a crude estimate, is within a factor of two of the purely theoretically-predicted value for monolayer CrCl$_{3}$ (300\,m/s). 
Furthermore, the anti-ferromagnetic coupling between the layers is expected to soften the magnon sound~\cite{essler2004haldane}, consistent with the monolayer sound velocity being larger than the bulk one.
A better quantitative understanding of the spectra requires a precise model of the coupling between the drive field and the magnetic modes of the sample (see also supplement XVIII).

Remarkably, the fitted linewidth of this resonance grows with decreasing temperature (Fig.~\ref{fig5}D). It is important to point out that due to the finite bandwidth of the NV momentum filtering, the linewidths of the spectra obtained in Fig.~\ref{fig5} cannot be employed as a direct measurement of the sound mode damping rate (see SI XVIII). Nevertheless, the observation of enhanced linewidth with decreasing temperature (Fig.~\ref{fig5}D) is notable, and could in part reflect the anomalous trend in viscous damping which we employed to explain the scaling of temperature-dependent noise observed in monolayer CrCl$_{3}$ (Fig.~\ref{fig3}D).

\section{Discussion}\label{sec12}

The presence of a hydrodynamic magnon sound mode in CrCl$_{3}$ provides a consistent picture of several complementary observations in the present work. In particular, the anomalous trend in the temperature-dependent noise generated by monolayer CrCl$_{3}$ indicates the presence of a viscously-damped hydrodynamic magnon sound mode (Fig.~\ref{fig3}A,D). Our spectroscopy measurements on multilayer CrCl$_{3}$ provide direct evidence for this mode (Fig.~\ref{fig5}C). 

The quantitative properties of a hydrodynamic magnon gas are set by two transport coefficients---the magnon sound velocity and viscosity. The present work reveals two exciting opportunities to explore these coefficients via both theoretical and experimental developments. Notably, the estimated experimental velocity of the sound mode (150\,m/s) is slower than that expected in our semi-classical theory (300\,m/s). Part of this discrepancy may reflect the crudeness of our experimental estimate, but it is also likely that the system exhibits physics beyond our present understanding. In particular, our prediction of the sound mode velocity is based on the dispersion of a bare magnon band. The presence of a hydrodynamic magnon sound mode, however, originates from strong interaction effects between the magnons, which are in turn expected to enhance the magnon effective mass~\cite{goldhirsch1980magnon} and slow down the sound mode relative to the results of our spin-wave theory. 

Furthermore, experimentally extracting the other key transport coefficient, viscosity $\mu$, from either the equilibrium noise measurements (Fig.~\ref{fig3}) or the driven experiments (Fig.~\ref{fig5}) requires precise knowledge of the magnon mass. Thus, a more complete theoretical picture of magnon renormalization effects and independent experimental measurement of the high-momentum magnon dispersion are critical for a quantitative understanding of the hydrodynamic magnon regime.

More broadly, this work serves as a starting point for exploring the physics of magnon sound in magnets~\cite{michel1970hydrodynamic,reiter1968magnon,rodriguez2022probing}, a platform with key advantages over the more familiar case of second-sound in superfluids~\cite{hohenberg1965microscopic}. For example, coherent magnon sound waves could be realized at ambient conditions in magnets with large exchange constants~\cite{buczek2020first}. Such longitudinal spin waves even exhibit properties that may find value in spintronic, magnonic, and caloritronic devices~\cite{ulloa2019nonlocal,sano2023breaking}.

\section{Acknowledgments}
The authors would like to thank I. Esterlis, J. Rodriguez-Nieva, M. Ziffer, and B. Halperin for fruitful discussions, N. R. Reed for support on the forward Fourier method, and M. Tschudin for assistance with the experiment. This work is supported by the Quantum Science Center (QSC), a National Quantum Information Science Research Center of the U.S. Department of Energy (DOE). A. Y. is also partly supported by the Gordon and Betty Moore Foundation through Grant GBMF 9468, and by the U.S. Army Research Office (ARO) MURI project under grant number W911NF-21-2-0147. R. X. is also partly supported by the Army Research Office under Grant numbers: W911NF-22-1-0248. N.M. is supported by an appointment to the Intelligence Community Postdoctoral Research Fellowship Program at Harvard University administered by Oak Ridge Institute for Science and Education. The work of P.E.D. is sponsored by the Army Research Office and was accomplished under Grant Number W911NF-21-1-0184. A.M. and E.D. are supported by the SNSF project 200021$\_$212899, the Swiss State Secretariat for Education, Research and Innovation (SERI) under contract number UeM019-1 and NCCR SPIN, a National Centre of Competence in Research, funded by the Swiss National Science Foundation (grant number 225153). F.M. acknowledges support from the NSF through a grant for ITAMP at Harvard University. 
PJH acknowledges supports by by the Army Research Office MURI (grant no. W911NF2120147), the 2DMAGIC MURI (grant no. FA9550-19-1-0390), the National Science Foundation (grant no. DMR-1809802), the Office of Naval Research (grant no. N000142412440), the Ramón Areces Foundation and the Gordon and Betty Moore Foundation’s EPiQS Initiative through grant no. GBMF9463.

\section*{Author contributions}

R.X., N.M., and A.Y. conceived the experiment. R.X., N.M., and R.K. performed the experiment and data analysis. P.E.D., A.M., and F.M. developed the theoretical interpretation of the experiment, with critical input from A.Y., E.D., and M.D.L. L-Q.X. and R.X. fabricated the samples under the supervision of P.J.-H.. D.R.K. and D.M. grew and characterized the CrCl$_3$ crystals under the supervision of P.J.-H.. K.W. and T.T. provided the hBN crystals. All authors contributed to the writing of the manuscript.

\section*{Declarations}

The authors declare no conflicts of interest. Raw data and analysis code are available upon reasonable request.

\bibliography{sn-bibliography}

\end{document}